\documentclass{llncs}

\usepackage[svgnames]{xcolor}
\usepackage{multicol}
\usepackage[utf8]{inputenc}
\usepackage{hyperref}
\usepackage{blindtext}
\usepackage{listings}
\usepackage{enumerate}
\usepackage{dirtytalk}
\usepackage{amsmath, amssymb, amsfonts}
\usepackage[export]{adjustbox}
\usepackage{tikz}
\usepackage{subcaption}
\usepackage{syntax}
\usepackage{xspace}
\usepackage{algorithm}
\usepackage{mathpartir}
\usepackage[noend]{algpseudocode}
\usepackage{wrapfig}
\usepackage{pstricks-add}
\usepackage{graphicx}
\usepackage{booktabs}
\usepackage{todonotes}
\usetikzlibrary{shapes, shapes.misc, positioning, calc, fit, arrows, decorations.pathreplacing}

\renewcommand{\subsubsection}[1]{\paragraph{#1}}

\newtheorem{hypothesis}{Hypothesis}

\newcommand{\kw}[1]{\textsf{#1}}
\newcommand{\assert}[1]{\ensuremath{\texttt{/*@ }\kw{assert}~#1\texttt{; */}}}
\newcommand{\letin}[2]{\kw{let}~#1 = #2\texttt{;}~}
\newcommand{\builtin}[1]{\textsf{#1}}
\newcommand{\bs}{\ensuremath{\backslash}}

\newcommand{\commentJS}[1]{{\color{blue}{\textbf{Julien:} #1}}}
\newcommand{\commentFV}[1]{{\color{orange}{\textbf{Franck:} #1}}}
\newcommand{\commentMJ}[1]{{\color{ForestGreen}{\textbf{Maxime:} #1}}}
\newcommand{\commentFM}[1]{{\color{magenta}{\textbf{Fonenantsoa:} #1}}}
\newcommand{\commentNK}[1]{{\color{violet}{\textbf{Nikolai:} #1}}}

\renewcommand{\commentJS}[1]{}
\renewcommand{\commentFV}[1]{}
\renewcommand{\commentMJ}[1]{}
\renewcommand{\commentFM}[1]{}
\renewcommand{\commentNK}[1]{}

\newcommand{\name}[1]{\textsf{#1}\xspace}
\newcommand{\fluctuat}{\name{Fluctuat}}
\newcommand{\framac}{\name{Frama-C}}
\newcommand{\acsl}{\name{ACSL}}
\newcommand{\eacsl}{\name{E-ACSL}}
\newcommand{\fldcompiler}{\name{FLDCompiler}}
\newcommand{\fldlib}{\name{FLDLib}}
\newcommand{\cadna}{\name{Cadna}}
\newcommand{\gmp}{\name{GMP}}
\newcommand{\Eva}{\name{Eva}}

\newcommand{\C}{\name{C}}
\newcommand{\Cpp}{\name{C++}}

\newcommand{\idfloat}{_{\mbox{\scriptsize\em float}}}
\newcommand{\idreal}{_{\mbox{\scriptsize\em real}}}
\newcommand{\iderror}{_{\mbox{\scriptsize\em err}}}
\newcommand{\Split}{\texttt{split}\xspace}
\newcommand{\Merge}{\texttt{merge}\xspace}

\newcommand{\mustdef}{\textit{mustdef}}
\newcommand{\maydef}{\textit{maydef}}
\newcommand{\mayref}{\textit{mayref}}
\newcommand{\data}{\textit{data}}

\newtheorem{S4Rrule}{Criterion}

\begin{document}


\title{Abstract Compilation for Verification of\\
  Numerical Accuracy Properties}
\author{Maxime Jacquemin$^{1}$
  \and Fonenantsoa Maurica$^{1,3}$
  \and Nikolai Kosmatov$^{1,2}$
  \and\\ Julien Signoles$^{1}$
  \and Franck V\'edrine$^{1}$}
\institute{
  $^1$CEA, LIST, Software Security and Reliability Laboratory, Palaiseau, France\\ 
  \email{firstname.lastname@cea.fr}\\
  $^2$Thales Research \& Technology, Palaiseau, France\\
  \email{nikolaikosmatov@gmail.com}\\
  $^3$Billee, Neuilly-sur-Seine, France\\
  \email{firstname.lastname@billee.fr}
}
\maketitle
\vspace{-5mm}
\begin{abstract}
Verification of numerical accuracy properties in modern software remains an important and challenging task.
This paper describes an original framework combining different solutions for numerical accuracy.
First, we extend an existing runtime assertion checker called \eacsl  with rational numbers to monitor accuracy properties at runtime.
Second, we present an abstract compiler, \fldcompiler, that performs a source-to-source transformation
such that the execution of the resulting program, called an \emph{abstract execution}, is an
abstract interpretation of the initial program.
Third, we propose an instrumentation library \fldlib that
formally propagates accuracy properties along an abstract 
execution. While each of these solutions has its own interest,
we emphasize the benefits of their combination for
an industrial setting.
Initial experiments 
show that the proposed technique can efficiently and soundly
analyze numerical accuracy
for industrial 
programs
on thin numerical scenarios.
\end{abstract}





\section{Introduction}

Floating-point numbers are widely used in many areas such as digital signal
processing, localization applications, neural networks, and high performance
computing. However, the results of floating-point operations are rounded, which
approximates them w.r.t.\ ideal computations on real numbers. An
accumulation of rounding errors may lead to inaccurate computations that are sometimes at the origin of disastrous bugs%
\footnote{\url{http://www-users.math.umn.edu/~arnold/disasters/patriot.html}}%
\footnote{\url{https://en.wikipedia.org/wiki/Vancouver\_Stock\_Exchange}}%
\footnote{\url{http://www-users.math.umn.edu/~arnold/disasters/sleipner.html}}.
Therefore, it is of the utmost importance to verify the \emph{accuracy} of such
numerical computations in critical systems.

Verification
of accuracy properties remains a challenging
research topic since many years~\cite{monniaux08toplas}. Testing-based
techniques are too optimistic: they do not detect many numerical misbehaviors
because they cannot represent real numbers precisely. \emph{A contrario}, static
analysis techniques, for instance, based on abstract
interpretation~\cite{cousot77popl}, are too pessimistic: they compute
over-approximations of the floating-point operations that lead to false
alarms, signaling potential errors that cannot happen on any concrete
execution.

In this paper, we propose the new concept of \emph{abstract compilation} as an
intermediate technique between testing and abstract interpretation
. An abstract compiler embeds an abstract
interpretation engine for floating-point computations into the generated code in
order to soundly verify the properties of interest during its execution, called
an \emph{abstract execution}. To that end, the abstract
compiler converts each floating-point value
of the concrete
execution to an abstract value 
that is interpreted at runtime by a dedicated library
.

The main difficulty of this technique consists in handling \emph{unstable tests} in
a sound way. Indeed, an unstable test happens when the guard of a conditional
depends on a floating-point expression $e$ and can be evaluated to a boolean
value different from the one relying on real arithmetic.
In such a case, the program's concrete execution flow differs from the theoretical one since the conditional branch that is executed is not the
same. To solve this issue, each unstable test is enclosed in
a loop in order to soundly iterate over all possible execution paths.

Compared to abstract interpretation, abstract compilation remains precise even if
the execution context is not statically known (e.g.\ in presence of
communication channels), since the generated code embedding the
abstract engine is executed on concrete inputs. Compared to testing techniques,
it remains sound. It is also possible to test programs while taking into account
uncertainty of their inputs (e.g. coming from sensors), providing better
insights on its robustness.

Last but not least, when the numerical accuracy properties of interest are
actually properties over \emph{rational numbers} (in $\mathbb{Q}$), the
generated code evaluates them in an exact manner, without relying on abstract
values. Therefore, no approximations are introduced in such cases.

For evaluating this approach in practice,
we have implemented a prototype abstract compiler from \C
code annotated with formal numerical accuracy properties
into \Cpp{} code.
It relies on three main
components: the runtime assertion checker \eacsl~\cite{signoles17:eacsl} to
convert the formal properties into \C code, \fldcompiler{} that deals with
instability and generates the resulting \Cpp{} code, and the \Cpp{}
library \fldlib~\footnote{\url{https://github.com/fvedrine/fldlib}} that
interprets abstract numerical values at runtime. This toolchain has been
evaluated on our own motivating examples, on small-size examples coming from
other sources, and on large-size industrial case studies (mostly synchronous
reactive systems of several dozens of thousands lines of code).
%
Each component of the toolchain has its own interest and has been developed
independently from abstract compilation. Hence, it is possible to use them
separately, or to replace one of these components by a different one in a particular
setting. For instance, it is possible to replace \fldlib{}
by \cadna~\cite{FJ-JMC-CPC-2008} to obtain an accuracy verification by
stochastic propagation.

To sum-up, the contributions of the paper include:
\begin{itemize}
\item
  an extension of the runtime assertion checker \eacsl{} that supports
  operations over rational numbers in an exact manner. As far as we know, that
  is the first support of rational numbers in such a tool;
\item
  a description of the abstract compiler \fldcompiler{} that deals with
  unstable tests while generating code for the abstract interpretation engine;
\item
  an
  instrumentation library, \fldlib, that formally propagates
    accuracy properties along the abstract execution of the test program;
\item
  a presentation of a new verification technique named \emph{abstract
  compilation} that embeds an abstract interpretation engine at runtime, as well
  as a prototype toolchain that combines the three aforementioned components;
\item
  an evaluation of this toolchain over a set of representative programs.
\end{itemize}

\paragraph{Outline.}
The rest of the paper is organized as follows. Section~\ref{sec:compil}
presents an overview of abstract compilation through a motivating
example. Section~\ref{sec:background} provides the necessary background in
order to understand the technical parts of the
paper. Section~\ref{sec:toolchain} details the three components of our abstract
compiler toolchain. Section~\ref{sec:experimental} shows our experimental
results. Section~\ref{sec:related-work} introduces the related work before
concluding and introducing perspectives in Section~\ref{sec:conclusion}.

\section{Overview and Motivating Example}
\label{sec:compil}

Abstract compilation is an intermediate verification technique for numerical
accuracy between testing and abstract
interpretation~\cite{cousot77popl}. Testing tools,
e.g. \name{FpDebug}~\cite{benz12pldi}, are likely to deliver too optimistic
results since they have no confidence interval, whereas abstract interpreters,
e.g.  \fluctuat~\cite{goubault11vmcai}, are likely to deliver too pessimistic
results (false alarms) that come from over-approximations.

Consider for instance the \C function of Fig.~\ref{fig:motiv-example}.
It implements an interpolation table \texttt{y} composed of \texttt{n}
measures to compute a linear approximation of a continuous function on a point
\texttt{in} $\in [0, n-1]$.
We would like to verify that, at every call site, the round-off error of the
result can increase the imprecision of the input by at most twice the biggest measure of the table.
We would also like to check that the function is robust enough: not only this
property is satisfied for all concrete input values, but also for any value near
them~\cite{goubault13aplas}.

For instance, in the \acsl specification language~\cite{baudin:acsl},
the desired accuracy and robustness properties can be
expressed  by the following assertion: 
\lstset{language=C,
	keywordstyle=\tt\bf\color{black},commentstyle=\it\color{blue}\ttfamily,
	showlines=true, moredelim=**[is][\color{gray}]{@}{@}, mathescape,
	basicstyle=\small\ttfamily}
\begin{lstlisting}
/*@ assert
      \let (err_min, err_max) = accuracy_get_ferr(in);
      \let cst = max_distance(&y[0], n);
      accuracy_assert_ferr(out,
        -2.0 * cst * min(err_min, -err_max),
        +2.0 * cst * max(-err_min, err_max)); */
\end{lstlisting}
Function \texttt{max\_distance} 
(omitted here)
computes the maximal distance between two successive
elements of \texttt{y}, 
that is, $\operatorname{max}_{i=0,\dots,n-2} | y[i+1] - y[i] |$. 
Functions \texttt{accuracy\_get\_ferr}
and \texttt{accuracy\_assert\_ferr} are built-ins that extend the
original \acsl language. The first one gives access to the error bounds of its
parameter while the second one asserts an accuracy property, here that the
resulting round-off error for \texttt{out} is bounded
by $-2.0 \times
\texttt{cst} \times \operatorname{min}(\texttt{err\_min},-\texttt{err\_max)}$
and
$2.0 \times \texttt{cst} \times \operatorname{max}(-\texttt{err_min}, \texttt{err_max})$. 

It is worth noting that the robustness property is actually broken for this example 
since for two close values
$-1$ and $-1+\varepsilon$ of \texttt{in} (with a small $\varepsilon > 0$), we have \texttt{index} equal, resp., to -1 and 0, and  
\texttt{out} equal, resp., to $\texttt{y[0]}$ and $2 \texttt{y[0]}- \texttt{y[1]}$, with 
an obvious discontinuity.
Therefore, an alarm shall be raised if (and, optimally, only if) such a
context is encountered.

\begin{figure}[tb]
\lstset{language=C,
keywordstyle=\tt\bf\color{black},commentstyle=\it\color{blue}\ttfamily,
showlines=true, moredelim=**[is][\color{gray}]{@}{@}, mathescape,
basicstyle=\small\ttfamily,numbers=left}
\begin{lstlisting}
double interpolate(double in, double *y, int n) {
  double out;
  int index = (int) in;
  if (index < 0 || index >= n-1)
    out = (index < 0) ? y[0] : y[n-1];
  else
    out = y[index] + (in - index) * (y[index+1] - y[index]);
  return out;
}
\end{lstlisting}
\vspace{-4mm}
\caption{Motivating example: an interpolation table.}
\label{fig:motiv-example}
\end{figure}

Testing-based techniques cannot check the desired 
properties for all values, while abstract interpreters have a hard time to verify them in a
precise manner because they include three verification challenges:
\begin{enumerate}
\item keeping precise relationships between variables \texttt{index}
and \texttt{in} after the initialization of the former through the latter at
line 3;
\item analyzing the unstable test issued from line 3 that may lead to execute
one branch of the conditional or another depending on the guard is computed over
floating-point or real numbers;
\item expressing the desired accuracy and robustness properties in a formal way.
\end{enumerate}
In addition to these technical challenges, a practical abstract interpreter
usually requires to stub input-output (I/O) functions such as 
communications with the environment and the initialization machinery.
To circumvent these issues, our approach embeds into the code an 
abstract interpretation engine that only manages  floating-point
computations. Concrete code execution solves the practical issues of
a static abstract interpreter. It also solves point (1) since the
relations are 
implicitly kept by the execution flow, while the abstract compiler
automatically replaces the concrete floating-point values and operators by their
abstract counterparts that soundly take into account round-off errors.
The abstract compiler also takes care of point (2) by enclosing each potential
unstable test in a loop, so ensuring a complete coverage of all possible
executions as explained in Section~\ref{sec:toolchain:fldcompiler}.
%
Finally, point (3) is addressed by relying on a formal specification language,
e.g. \acsl~\cite{baudin:acsl}, that can express powerful numerical
properties. 

\section{Background}
\label{sec:background}
This section introduces the necessary background for understanding the technical
parts of the paper: \textsc{Ieee}-754 floating-point numbers in
Section~\ref{sec:float}, and the minimal required specification language based
on \acsl in Section~\ref{sec:acsl}.

\subsection{
IEEE-754 Floating-Point Numbers}
\label{sec:float}

A floating-point number is a rounded representation of a real number.
Simply put, a real number $x$, denoted as
$x = {(-1)}^s m \beta^e$ in scientific notation, is approximated by the
floating-point number
$f = \circ(x) = {(-1)}^s \hat{m} \beta^e$ where $s \in \{0,1\}$,
$\beta \in \mathbb{N}$, $\beta \geq 2$,
$e \in \{e_{\min},\dots, e_{\max}\}\subset \mathbb{Z}$,
$m \in \mathbb{R}^+$, and $\hat{m}$ is an approximation of $m$ on $p \geq 2$
digits.
The quantities $\beta, p, e_{\min}$ and $e_{\max}$ parameterize the considered
floating-point type.
The function $\circ$ is called the \textit{rounding function}.

Floating-point numbers are standardized by the \textsc{Ieee-$754$} norm. Among
others, it defines several standard types --- several assignments for
$\beta$, $p$, $e_{\min}$ and $e_{\max}$ --- such as the \texttt{float}
and \texttt{double} types of the \C programming language.
\textsc{Ieee}-754 defines several rounding functions, but the rest of the paper
assumes that $\circ$ rounds $x$ to the \emph{nearest} floating-point number.
A \textsc{Ieee}-754 compliant implementation also requires
floating-point arithmetic operations to be \emph{correctly rounded}: given a
real arithmetic operation $\star$, its floating-point equivalent
$\textcircled{$\star$}$ must satisfy $f_1 \textcircled{$\star$} f_2 = \circ(f_1
\star f_2)$ where $f_1, f_2$ are floating-point numbers.

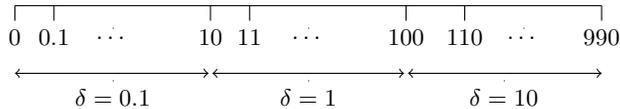
\begin{figure}[tb]
\centering
\begin{tikzpicture}
\def\x{0.26}

\draw (0*\x,1) -- (30*\x,1);

\draw (0*\x,1) -- (0*\x,0.8) node[anchor=north] {0};
\draw (2*\x,1) -- (2*\x,0.8) node[anchor=north] {0.1};
\draw(5*\x,0.7) -- (5*\x,0.7) node[anchor=north] {$\dots$};
\draw[<->] (0*\x,0.1) -- (9.9*\x,0.1) node[anchor=south] {};
\draw(5*\x,0) -- (5*\x,0) node[anchor=north] {$\delta = 0.1$};

\draw (10*\x,1) -- (10*\x,0.8) node[anchor=north] {10};
\draw (12*\x,1) -- (12*\x,0.8) node[anchor=north] {11};
\draw(15*\x,0.7) -- (15*\x,0.7) node[anchor=north] {$\dots$};
\draw[<->] (10.1*\x,0.1) -- (19.9*\x,0.1) node[anchor=north] {};
\draw(15*\x,0) -- (15*\x,0) node[anchor=north] {$\delta = 1$};

\draw (20*\x,1) -- (20*\x,0.8) node[anchor=north] {100};
\draw (23*\x,1) -- (23*\x,0.8) node[anchor=north] {110};
\draw(26*\x,0.7) -- (26*\x,0.7) node[anchor=north] {$\dots$};
\draw[<->] (20.1*\x,0.1) -- (30*\x,0.1) node[anchor=north]{};
\draw(25*\x,0) -- (25*\x,0) node[anchor=north] {$\delta = 10$};

\draw (30*\x,1) -- (30*\x,0.8) node[anchor=north] {990};





\end{tikzpicture}
\caption{%
Schematic representation of 
positive values of a floating-point type with $\beta=10$, $p=2$, $e_{\min}=0$ and $e_{\max}=2$. The difference between two consecutive floats, denoted by $\delta$, is different for different intervals of values.}%
\label{fig:myfloat}
\end{figure}

To illustrate all this, consider the floating-point type represented in
Fig.~\ref{fig:myfloat} and parameterized by $\beta = 10$, $p = 2$, $e_{\min} =
0$ and $e_{\max} = 2$.
In this type, the real $x = 10\pi = {(-1)}^0 3.14\cdots 10^{1}$ is
approximated by $f = o(x) = {(-1)}^0 3.1 \cdot 10^{1} = 31$ while computing
$1/3$ gives us $1 \textcircled{$/$} 3 = \circ(1/3) = \circ(0.3\cdots) = 0.3$.
In both examples, the rounding introduces a difference
between the behavior of the real numbers and the floating-point numbers.
This difference, called the \emph{absolute error}, is normally quite small with
standard floating-point type.
Its counterpart, the \emph{relative error}, is defined as the absolute error
divided by the expected result in real numbers.
Since these errors are accumulated
at each floating-point operation of a program, they may lead to huge
differences between the observed and expected behaviors.


The interested reader may refer to~\cite{Goldberg:1991:What}
or~\cite{Muller:2010:Ulp} for additional details.

\subsection{Specification Language}
\label{sec:acsl}


The purpose of this work is to target accuracy properties of \C programs. For
expressing them, we rely on the \eacsl specification
language~\cite{signoles:eacsl}, derived from 
the \acsl specification language~\cite{baudin:acsl}. 
The differences between both languages~\cite{delahaye13sac} are of no importance
for our work, so we only present here an overview of a small common
fragment. Its formal syntax is shown in Fig.~\ref{fig:syntax}.
Most constructs have already been illustrated in Section~\ref{sec:compil}.

\begin{figure}[tb]
\begin{align*}
assert & ::= \assert{pred} & \mbox{assertion} \\
pred &::= pred~rel~pred & rel \in \{ \wedge, \vee, \implies \} \\
& | \quad \neg pred & \mbox{negation} \\
& | \quad term~cmp~term & cmp \in \{ <, \le, \equiv, >, \ge \} \\
& | \quad \letin{x}{term} pred & \mbox{local binding} \\
& | \quad p(term, \ldots, term) & p \in Pbuiltins\\
term &::= x & \mbox{logic binder} \\
& | \quad lv & \mbox{left-values}\\
& | \quad zcst & zcst \in \mathbb{Z} \\
& | \quad qcst & qcst \in \mathbb{Q} \setminus \mathbb{Z} \\
& | \quad term~op~term & op \in \{ +, -, \times, / \}
\end{align*}
\vspace{-5mm}
\caption{Syntax of the specification language.}\label{fig:syntax}
\end{figure}

Logic statements are assertions enclosed in special comments \texttt{/*@ \ldots
  */} that may be written before any \C instruction. Assertions are typed
predicates which include 
logical relations and comparison operators
over terms, local bindings \emph{\`a la} \name{ML}, as well as applications of
built-in predicates shown in Fig.~\ref{fig:builtins}, in which $\mathbb{F}$
denotes either type \texttt{float} (if \texttt{f}) or \texttt{double}
(if \texttt{d}).
These predicates have their counterparts supported by the \fldlib library
(cf. Section~\ref{sec:toolchain:fldlib}). The pair of
built-ins starting with \texttt{accuracy\_enlarge} enlarge the values and the
absolute errors to the two pairs of bounds provided as
arguments. \texttt{accuracy\_assert} built-ins check whether the absolute or (if \texttt{rel} is indicated)
the relative error is included within the given bounds. \texttt{accuracy\_get}
built-ins return the lower and upper bounds of the absolute or
relative  error, or those of the real-number domain. The last built-ins print
information about the internal \fldlib representation.

\begin{figure}[tb]
  \centering
  \begin{tabular}{|l@{:~}l|}
    \hline
    Built-in name & type \\
    \hline
    \builtin{accuracy\_enlarge\_[f,d]val\_err} %
        & $\mathbb{F}
        \times \mathbb{Q}^4 \rightarrow \texttt{bool}$ \\
     \builtin{accuracy\_assert\_[f,d][rel]err}
        & $\mathbb{F}
        \times \mathbb{Q}^2 \rightarrow \texttt{bool}$ \\
     \builtin{accuracy\_get\_[f,d][rel]err}
        & $\mathbb{F} \rightarrow \mathbb{Q}^2$ \\
     \builtin{accuracy\_get\_[f,d]real}
        & $\mathbb{F} \rightarrow \mathbb{Q}^2$ \\
     \builtin{[f,d]print}
        & $\mathbb{F} \rightarrow \texttt{bool}$ \\
    \hline
    \end{tabular}
    \caption{Built-in predicates extending \acsl. Their counterparts exist in  \fldlib.}\label{fig:builtins}
\end{figure}

Terms are logic binders $x$, \C left-values $lv$ (variables, pointer derefences,
array and struct accesses, etc), mathematical constants (either integers or
rationals), or numerical operators. Terms are typed. The typing rules are left
implicit here, but are straightforward. A numerical operation is an integer one
if both arguments are integers; otherwise it is an operation over rational
numbers (and the integer argument, if any, is automatically promoted to the
corresponding rational number).
It is worth noting that all constants and numerical operators are over
mathematical numbers (either integers in $\mathbb{Z}$, or rationals in
$\mathbb{Q}$ depending on the context). \C integers and floating-point values are
implicitely coerced to their mathematical counterparts. For the sake of
simplicity, we assume no \texttt{NaN} nor $\pm \infty$ values, as well as no
runtime errors when evaluating \C left-values (see~\cite{delahaye13sac} for a
discussion about them).
%


%
\section{Abstract Compilation Toolchain}
\label{sec:toolchain}
\begin{figure}[tb]
\centering
\begin{tikzpicture}
  \tikzstyle{prgm} = [draw, rectangle, minimum size=4mm]
  \node (annot) at (0,1.45) [prgm] { annotated \C code };
  \node (instr) at (2.8,0) [prgm] { instrumented \C code };
  \node (cpp) at (5.6,1.45) [prgm] { \Cpp code with unstable tests };
  \node (bin) at (9,0) [prgm] { instrumented binary };
  \tikzstyle{compil} = [midway, sloped]
  \tikzstyle{edge} = [->, >=latex, draw=ForestGreen, line width = 1.4pt]
  \tikzstyle{edge} = [->, >=latex, draw, line width = 0.5pt]
  \draw[edge] (annot) -- (instr) node[compil,above] { {{\scriptsize ~\eacsl}} };
  \draw[edge] (instr) -- (cpp) node[compil,above]
              {\hspace{-4mm}{{\scriptsize \fldcompiler}}};
  \draw[edge] (cpp) -- (bin) node[compil,above] { {{\scriptsize\fldlib}} };
  \draw[edge] (cpp) -- (bin) node[compil,below] { \scriptsize or {{\cadna}} };
\end{tikzpicture}
\vspace{-2mm}
\caption{The abstract compilation toolchain.}\label{fig:toolchain}
\end{figure}
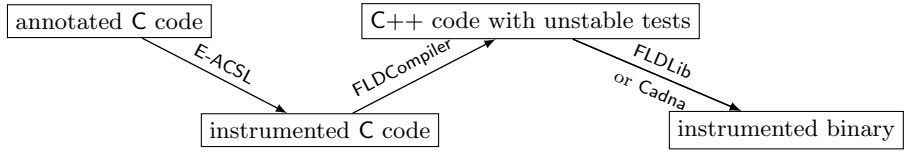

Our toolchain, shown in Fig.~\ref{fig:toolchain}, consists of three
components, 
respectively introduced in
Sections~\ref{sec:toolchain:eacsl},~\ref{sec:toolchain:fldcompiler},
and~\ref{sec:toolchain:fldlib}. 
The first two steps, \eacsl and \fldcompiler, perform
code generation,
while the last step 
is linking with \fldlib library (that can be replaced
at no cost by \cadna~\cite{FJ-JMC-CPC-2008}).
Steps 2, 3 are further detailed in Fig.~\ref{fig:fldlib-architecture}.
%

\subsection{E-ACSL Runtime Assertion Checker}
\label{sec:toolchain:eacsl}


\eacsl~\cite{signoles17:eacsl} is the runtime assertion checker of
\framac~\cite{kirchner15:framac}, an analysis framework for \C code. It
converts a \C program annotated with \acsl formal specifications to another \C
program that behaves at runtime as the original one if every annotation is
satisfied, otherwise it fails on the first invalid annotation. Therefore, it may be
seen as a compiler from \acsl\!\!+\C code to \C code.
While this tool has been being developed since a few years, it has been extended here to
support numerical properties over rational numbers\footnote{This part of the
  work will be included in the next open source version of \framac
  .}. In addition, a special extension to support the numerical
built-ins presented in Fig.~\ref{fig:builtins} has also been designed for our
abstract 
compiler. It generates \fldlib-compatible code.

As explained in Section~\ref{sec:acsl}, the specification language relies on
mathematical integers and rational numbers. Therefore, the generated code cannot
soundly use standard \C operators over integral or floating-point
types. Instead, \eacsl generates special code relying 
on \gmp library\footnote{\url{https://gmplib.org/}}
to soundly represent mathematical integers and rationals. Built-in predicates of Fig. \ref{fig:builtins}
are directly compiled into their \fldlib counterparts. 

However, the generated \gmp code
is inefficient compared to machine instructions. 
To reduce the need for \gmp code without loss of soundness, we have
designed a dedicated type system in order to  rely on machine representations
as much as possible, and generate \gmp code only when necessary.
The type system relies on integer intervals and a
custom notion of kind. \eacsl infers such a kind for each term,
and uses it to decide whether the machine types and operations 
can be sufficient to perform the required operations or not.
Our technical presentation
of \eacsl focuses on this type system.
In practice, thanks to this solution, very few
pieces of \gmp code are generated for integers.
Some comparisons of floating-point values
without operators (e.g. \texttt{f == 0.}) do not require \gmp's rational
operations either (as in the example in the end of the section).

We assume the existence of a type system, where $\Sigma(t)$ denotes the type of
a term $t$ and the primitive $\operatorname{isinteger}(t)$ (resp.
$\operatorname{isfloat}(t)$) is true if and only if $\Sigma(t)$ is a subtype of
$\mathbb{Z}$ (resp. a floating-point type). $\preccurlyeq_\tau$ is the subtyping
relation (expressing that all values of one type are also values of the other).

\paragraph{Integer intervals.}

We consider integer intervals with 
partial order $\preccurlyeq_I$.
Let $\mathbb{T}(I)$ be the smallest \C
integral type containing interval $I$, or $\mathbb{Z}$ otherwise, and $\mathbb{I}(t)$ be
an interval that contains all the possible values of the term $t$. In practice,
\eacsl relies on a simple syntactic type-based inference system to compute
$\mathbb{I}(t)$\footnote{This inference system is omitted here, but has already
  been presented 
  in French~\cite{jakobsson15jfla}.
  }.

\paragraph{Kinds.}
A \emph{kind} is either an integer interval, or a floating-point type, or a
rational. More formally, let $(\mathbb{K}, \preccurlyeq)$ be the
(sup-)semi-lattice of kinds defined as follows:

\vspace{-3mm}
\begin{minipage}[l]{.4\textwidth}			
\begin{align*}
  \mathbb{K} & ::= \mathcal{Z}~I & \mbox{integer interval } I \\
  & | \quad \mathcal{F}~\tau & \mbox{floating-point type } \tau \\
  & | \quad \mathcal{Q} & \mbox{rational}
\end{align*}
\end{minipage}
\hspace{15mm}
\begin{minipage}[l]{.4\textwidth}			
\begin{align*}
  \mathcal{Z}~I_1 \preccurlyeq \mathcal{Z}~I_2 &\iff I_1 \preccurlyeq_I I_2 \\
  \mathcal{F}~\tau_1 \preccurlyeq \mathcal{F}~\tau_2
  &\iff \tau_1 \preccurlyeq_\tau \tau_2 \\
  \mathcal{Z}~I \preccurlyeq \mathcal{F}~\tau
  &\iff \mathbb{T}(I) \preccurlyeq_\tau \tau
 \\
  K \preccurlyeq \mathcal{Q} & \quad (\mbox{for all } K \in \mathbb{K})
\end{align*}
\end{minipage}
	
Let $\cup$ denote the union over kinds induced by their lattice structure. The
kind of a term $t$, denoted $\kappa(t)$, and the type of a kind $k$,
denoted $\theta(k)$, are defined as follows:

\vspace{-3mm}
\begin{minipage}[l]{.4\textwidth}			
\begin{align*}
  \kappa(t) &= \mathcal{Z}~\mathbb{I}(t)
  & \mbox{if } \operatorname{isinteger}(t) \\
  \kappa(t) &= \mathcal{F}~\Sigma(t)
  & \mbox{if } \operatorname{isfloat}(t) \\
  \kappa(t) &= \mathcal{Q}
  & \mbox{if } \neg \operatorname{isfloat}(t)
\end{align*}
\end{minipage}
\hspace{15mm}
\begin{minipage}[l]{.4\textwidth}			
\begin{align*}
  \theta(\mathcal{Z}~I) &= \mathbb{T}(I) \\
  \theta(\mathcal{F}~\tau) &= \tau \\
  \theta(\mathcal{Q}) &= \mathbb{Q}
\end{align*}
\end{minipage}

\paragraph{Type system.}

Fig.~\ref{fig:typing} presents the type system. A type judgement, written
$\Gamma \vdash t: \tau_1 \leadsto \tau_2$ for terms (resp. $\Gamma \vdash_p p:
\tau_1 \leadsto \tau_2$ for predicates), means ``in the typing environment
$\Gamma$, the \C expression generated for $t$ (resp. $p$) may soundly have type
$\tau_1$, but, in the case of an operator (resp. a comparison), it must be
computed over type $\tau_2$.''. $\tau_2$ is omitted when irrelevant.  Predicates
return an \texttt{int}. For instance, assuming a large integer constant $c$ and
variables $x$ and $y$ of type \kw{int}, the term $x/(y+c)$ requires \gmp code
because $y+c$ does not fit into any \C type. However, its result fits into
a \kw{int}, so it may safely be compared to 0 with the usual \C
equality. Therefore, its type is
$\texttt{int} \leadsto \mathbb{Z}$. Fig~\ref{fig:deriv-tree} details the
derivation tree of $x/(y+c) \equiv 0$.
To improve precision, typing of
operators computes the kind of its operands and its result, merges them (to get
the most precise interval containing all of their possible values) and converts
the result into the corresponding \C type. The last two rules for terms are
coercion rules. The first one is a standard subsumption
rule~\cite{pierce02types}, while the second one soundly downcasts a term to a
smaller type than its own type if its infered kind fits in.

\begin{figure}[tb]
$$
\frac{}
     {\Gamma \vdash cst: \Sigma(cst)}
\qquad
\frac{}
     {\Gamma \vdash lv: \Sigma(lv)}
\qquad
\frac{}
     {\Gamma \vdash x: \Gamma(x)}
$$
$$
\frac{\tau = \theta(\kappa(t_1) \cup \kappa(t_2) \cup \kappa(t_1~\kw{op}~t_2))
  \quad \Gamma \vdash t_1: \tau
  \quad \Gamma \vdash t_2: \tau}
     {\Gamma \vdash t_1~\kw{op}~t_2: \tau \leadsto \tau}
$$
$$
\frac{\Gamma \vdash t: \tau' \quad \tau' \preccurlyeq_\tau \tau}
     {\Gamma \vdash t: \tau}
\qquad
\frac{\Gamma \vdash t: \tau' \leadsto \tau'
  \quad \tau \prec \tau'
  \quad \theta(\kappa(t)) \preccurlyeq_\tau \tau}
     {\Gamma \vdash t: \tau \leadsto \tau'}
$$
\bigskip
$$
\frac{\Gamma \vdash_p p_1: \kw{int} \quad \vdash_p p_2: \kw{int}}
     {\Gamma \vdash_p p_1~rel~p_2: \kw{int}}
\qquad
\frac{\Gamma \vdash_p p: \kw{int}}{\Gamma \vdash_p \neg p: \kw{int}}
\qquad
\frac{\Gamma \vdash t_1: \tau_1 \quad \cdots \quad \Gamma \vdash t_n: \tau_n}
  {\Gamma \vdash_p p(t_1, \dots, t_n): \kw{int}}
$$
$$
\frac{\tau = \theta(\kappa(t_1) \cup \kappa(t_2))
  \quad \Gamma \vdash t_1: \tau
  \quad \Gamma \vdash t_2: \tau}
     {\Gamma \vdash_p t_1~cmp~t_2: \kw{int} \leadsto \tau}
\qquad
\frac{\Gamma \vdash t: \tau \quad \Gamma, x: \tau \vdash_p p: \kw{int}}
{\Gamma \vdash_p \letin{x}{t} p: \kw{int}}
$$
\vspace{-2mm}
\caption{\eacsl type system.}\label{fig:typing}
\end{figure}

\begin{figure}[tb]
\vspace{-3mm}
$$
\inferrule* 
  {\inferrule* 
   {\inferrule* 
     {\vdash x: \kw{int}
     \\
     \inferrule* 
       {\vdash y: \kw{int} \\ \vdash c: \mathbb{Z}}
       {\vdash{y+c: \mathbb{Z}}}}
     {\vdash x/y+c: \mathbb{Z} \leadsto \mathbb{Z}}
     \\
     \kw{int} \prec \mathbb{Z}
     \\
     \theta(\kappa(x/(y+c)))\preccurlyeq \kw{int}}
   {\vdash x/y+c: \kw{int} \leadsto \mathbb{Z}}
  \\
  \vdash 0: \kw{int}
  }
  {\vdash_p x/(y+c) \equiv 0: \kw{int}}
$$
\vspace{-5mm}
\caption{Derivation tree for $x/(y+c) \equiv 0$.}
\label{fig:deriv-tree}
\end{figure}

\paragraph{Code generation.}

Generating code from the information computed by the type system is quite
straightforward. For instance, the code generated for the assertion
\assert{x/(y+c) \equiv 0 \wedge f-0.1 \leq g} with the first operand of $\equiv$
of type $\texttt{int} \leadsto \mathbb{Z}$, and $f$ and $g$ of type
\texttt{double} would be as follows
. 

\lstset{basicstyle=\scriptsize\ttfamily}
\begin{lstlisting}
/* compute x/(y+c) with GMP integers */
mpz_t _x, _y, _c, _add, _div; int _div2, _and;
mpz_init_set_si(_x, x); mpz_init_set_si(_y, y);
mpz_init_set_str(_c,"79228162514264337593543950335",10); /* some large value */
mpz_init(_add); mpz_add(_add, _y, _c);
mpz_init(_div); mpz_tdiv_q(_div, _x, _add);
/* safely downcast the result of the division from GMP to int for testing it */
_div2 = mpz_get_si(_div);
if (_div2 == 0) {
  /* compute f-0.1 <= g with GMP rationals */
  mpq_t _f, _cst, _g, _sub; int _le;
  mpq_init(_cst); mpq_set_str(_cst,"01/10",10);
  mpq_init(_f); mpq_set_d(_f, f);
  mpq_init(_sub); mpq_sub(_sub, _f, _cst);
  mpq_init(_g); mpq_set_d(_g, g);
  /* getting the result of the predicate as an int */
  _le = mpq_cmp(_sub, _g);
  _and = _le <= 0;
  /* de-allocate the allocated GMP variables for rationals */
  mpq_clear(_cst); mpq_clear(_f); mpq_clear(_sub); mpq_clear(_g);
} else
_and = 0;
/* runtime check the conjunction */
assert(_and);
/* de-allocate the allocated GMP variables for integers */
mpz_clear(_x); mpz_clear(_y); mpz_clear(_c); mpz_clear(_add);
mpz_clear(_div);
\end{lstlisting}

\subsection{\fldcompiler{} Abstract Compiler}
\label{sec:toolchain:fldcompiler}


As said previously, in order to be
sound, the embedding of an abstract interpretation engine inside the
program encloses each unstable test $t$ within a loop that must consider all possible evaluations of $t$.
The toolchain provides two directives to delimit those loops: \Split{} marks the
start of a block of code $B$ that must be run multiple times to analyze all 
possible executions, while \Merge{} marks the point of convergence where all 
memory states after the executions of $B$ must be joined into a unique state. Such a
block $B$ enclosed between these directives is called a \emph{split-merge
section}. 
They are inserted into the code by \fldcompiler
while their instrumentation is provided by \fldlib.
In the general case, \Split{} is parameterized by the variables that must be
reset before a new execution in order to ensure that the memory state is the same
at each loop iteration (i.e. each execution of the section runs from the same state), while \Merge{} is parameterized by the variables to be joined
after different executions. 

Consider again the example of Fig~\ref{fig:motiv-example}. 
\fldcompiler inserts a \Split{} directive
with no argument before the cast 
at line~3, while a \Merge{} directive parameterized by \texttt{out} is inserted
before 
line 8.
Indeed, a cast from a floating-point value to an integer is actually a form of an unstable
test since the real value can be cast to a different integer than the
floating-point one. 
The \Merge{} directive cannot be placed earlier in the function
because \texttt{out} would not be computed yet.

\begin{figure}[tb]
  \centering
  \scalebox{.8}{\includegraphics[width=\textwidth]{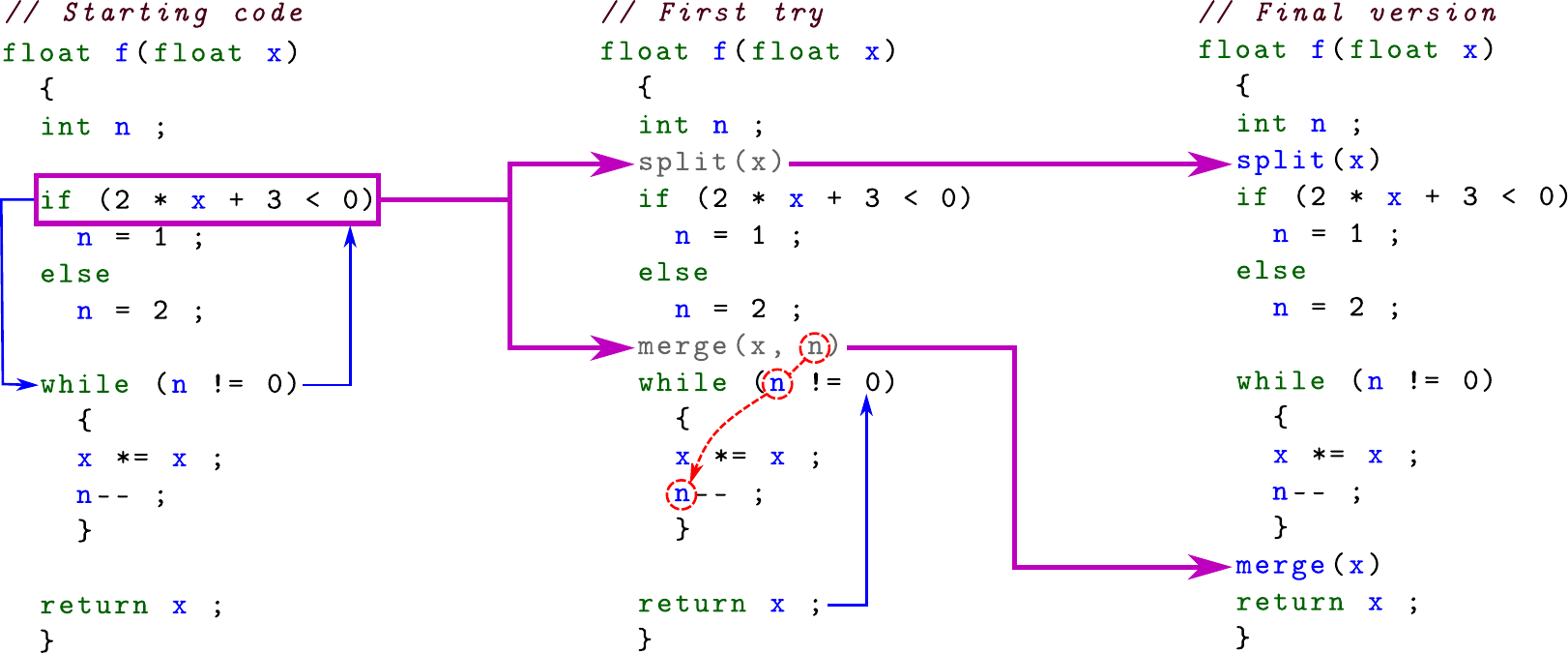}}
  \vspace{-3mm}
  \caption{Example of the transformation steps performed by \fldcompiler.}%
  \label{s4r/example}
\end{figure}

\fldcompiler{} is a source-to-source program transformation that automatically
annotates a program with the needed \Split{} and \Merge{} directives together
with their parameters. For the sake of performance and precision, a
generated split-merge section should be minimal, i.e. its block of
code is
as small as possible,  \Split{} only
resets what is 
needed, and \Merge{} only joins variables that are modified by the block of code
and used afterward. It is worth noting that computing these parameters
statically is undecidable, so over-approximations may actually be performed.
Positioning the split-merge sections is done by a greedy algorithm
that expands them through the code until three criteria, presented below, are
satisfied. These criteria are illustrated on the example of
Fig.~\ref{s4r/example} that contains the unstable test \texttt{if 
(2 * x + 3 < 0)}.

\begin{S4Rrule}
  A \Split{} must \emph{strictly dominate} its associated \Merge{}. Conversely,
  a \Merge{} must \emph{strictly post-dominate} its associated \Split{}.
\end{S4Rrule}

Dominance and post-dominance relations~\cite{Prosser:1959:ABM:1460299.1460314}
used in this criterion state that all paths that go through \Split{} must
go through its associated \Merge{} and, conversely, all paths that go through
\Merge{} must have gone through its associated \Split{}.
This criterion ensures that the memory allocations performed by \Split{} are
eventually freed by \Merge{}. 
The other way round, the memory freed by \Merge{} must have been initially
allocated by \Split{}.
In our example, the \texttt{if}
statement is \emph{post-dominated} by the \texttt{while}, which is
\emph{dominated} by the \texttt{if}.
Therefore, a \Split{} (resp. \Merge{}) directive is added before the \texttt{if}
(resp. \texttt{while}).

\begin{S4Rrule}
%
A split-merge section must start and end in the same block. 
%
\end{S4Rrule}

\begin{figure}[tb]
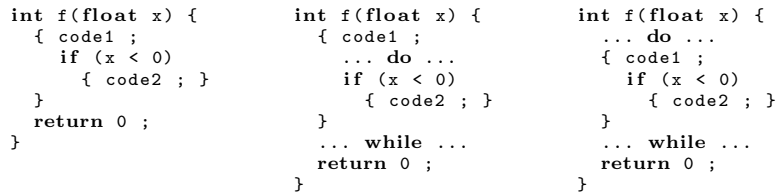

\centering
\begin{subfigure}[b]{0.3\textwidth}
  \begin{lstlisting}
   int f(float x) {
     { code1 ;
       if (x < 0)
         { code2 ; }
     }
     return 0 ;
   }


  \end{lstlisting}
\end{subfigure}
\begin{subfigure}[b]{0.3\textwidth}
  \begin{lstlisting}
   int f(float x) {
     { code1 ;
       ... do ...
       if (x < 0)
         { code2 ; }
     }
     ... while ...
     return 0 ;
   }
  \end{lstlisting}
\end{subfigure}
\begin{subfigure}[b]{0.3\textwidth}
  \begin{lstlisting}
   int f(float x) {
     ... do ...
     { code1 ;
       if (x < 0)
         { code2 ; }
     }
     ... while ...
     return 0 ;
   }
   \end{lstlisting}
\end{subfigure}
\vspace{-5mm}
\caption{(a) Original code, (b) invalid and (c) valid examples of instrumentation.}
\label{fig:criterion2-ex}
\end{figure}

A split-merge section is enclosed in a \texttt{do ... while} loop.
Its \texttt{do} part is generated by \Split{}, whereas the \texttt{while} part
is generated by \Merge{}. Generating them in distinct blocks would lead to
syntactically invalid \C code. In our example and most of the time, that is not
an issue, but enforcing it on nested blocks as in
Fig.~\ref{fig:criterion2-ex}a is trickier.
Here, Criterion~1 states that \Split{} must be inserted before \texttt{if},
while \Merge{} must be inserted before \texttt{return}. Inserting them
naively as in Fig.~\ref{fig:criterion2-ex}b leads to an invalid
code. \fldcompiler inserts them as in Fig.~\ref{fig:criterion2-ex}c.

\begin{S4Rrule}
  Non floating-point variables must be kept unchanged in every memory state
  generated by a \Split{} and joined by its associated \Merge{}.
\end{S4Rrule}

This criterion is mandatory because the libraries linked to the
generated code, e.g. \fldlib, 
have no abstraction for non floating-point variables: merging
them would result in a runtime crash.
For example, the middle part of Fig.~\ref{s4r/example} presents a first
positioning attempt for the split-merge section that actually violates
Criterion 3. Indeed, because the value of
the \emph{integer} variable \texttt{n} is modified in the \texttt{if} and is
needed after the \Merge{}, its values must be joined.
To fix this, \Merge{} is delayed as shown on the right side of
Fig.~\ref{s4r/example}.

\paragraph{Arguments of \Split{} and \Merge{}.}

As said previously, \Split{} and \Merge{} take parameters that specify, resp.,  the
variables to reset before a new execution of the section, and the ones 
to be eventually merged after them.
To minimize the analysis cost, only necessary parameters should be generated.
For example, if a variable is never modified, resetting its value is
useless. These parameters for \Split{} and \Merge{} are respectively computed by
a \emph{save-list} and a \emph{merge-list} whose computation is explained
below. They are based on a dedicated data dependency analysis inspired
from~\cite{DBLP:journals/fac/LechenetKG18}.
%
%
More precisely, for each statement $p$, this analysis gathers four sets,
informally defined as follows:
\begin{description}
  \item[$\mustdef(p)$]: a set of variables necessarily modified in $p$ (that is,
    all executions modify them). For instance, variable \texttt{n} of
    Fig.~\ref{s4r/example} is in the \mustdef{} set of \texttt{if}.

  \item[$\maydef(p)$]: a set of pairs $(x,s)$ where $s$ is a sub-statement of
    $p$ that may modify the variable $x$. 
    %
    In Fig.~\ref{s4r/example}, the \maydef{} set of
    \texttt{while} contains \texttt{x} and the loop body $b$ since executing $b$
    modifies \texttt{x}. However, \texttt{x} does not belong to the \mustdef{}
    set of \texttt{while} because, if \texttt{n} $= 0$, then \texttt{x} is left
    unchanged.

\item[$\mayref(p)$]: a set of pairs $(x,s)$ where $s$ is a statement of $p$
    that may read the variable $x$. 
    In Fig.~\ref{s4r/example}, \texttt{x} belongs to the \mayref{}
    set of \texttt{if} because it is read by its condition.
    For sequence of statements $S$, this set does not contain variables that are read after being assigned in $S$. For instance, \texttt{x} (paired to any statement)
    does not belong to \mayref{} of sequence $S = \texttt{x = 2; y = x + 3;}$.
%

\item[$\data(p)$]: a
    set of tuples $(s_1,s_2,x)$ in which $s_1$ writes a variable $x$ that is
    later read by $s_2$ (without intermediate writings).
    Its computations uses the three previous sets.
    For the example of sequence $S$ above,
    variable \texttt{x} is modified by \texttt{x = 2;} and then read by
    \texttt{y = x + 3}, so $(\texttt{x = 2}, \texttt{y = x + 3}, \texttt{x}) \in
    \data(S)$.    
\end{description}
The \textit{save-list} and \textit{merge-list} of a split-merge
section are computed as follows:
{%
\small
\begin{align*}
  \textit{save-list}(p) &= \{x \mid \exists (s_1,s_2),(x,s_1) \in \maydef(p)
  \wedge (x,s_2) \in \mayref(p)\}, \\
  \textit{merge-list}(p) &= \{x \mid \exists (s_1, s_2), (x,s_1) \in \maydef(p)
  \wedge (s_1, s_2) \in \data(\mathcal{F}(p)) \wedge s_2 \not\in p\},
\end{align*}
\vspace{-7mm}
\begin{align*}
  &\mbox{where } \mathcal{F}(p) \mbox{\ is the body of the function containing } p.
\end{align*}
\vspace{-4mm}
}

\noindent
It means that a variable $x$ is added to the \textit{save-list} of a section $p$
if there is a statement inside $p$ that may modify $x$ and another statement
that may read $x$.
Said another way, if a new execution may depend on the value of a variable that
could have been modified in another execution, then we need to reset it
before each execution.
Dualy, a variable $x$ is added to the \textit{merge-list} of a
section $p$ if there is a statement in $p$ that may modify $x$ and there is
another statement outside the section that may read $x$.

\fldcompiler{} is implemented as a \framac plugin and relies on its kernel to
pretty-print the generated code.
It visits the whole source code and generates the split-merge sections
based on the declared type of variables.  The basic version has no notion of
alias, so if a pointer iterates on the cells of a floating-point array, it does
not add them to the \textit{save-list} and the \textit{merge-list}, which may
produce unsound results. To get around this problem, \fldcompiler{} relies on
\name{Eva}~\cite{BlazyBYVMCAI2017}, the value analysis of \framac 
(as shown below in Fig.~\ref{fig:fldlib-architecture}), in order to
know all possible targets of pointers to be added to the 
\textit{save-list} and the \textit{merge-list}.
It is worth noting that it may add unnecessary variables to the lists since the
analysis by abstract interpretation is conservative.  Finally, \fldcompiler{}
issues a warning if it tries to add to the lists something that is dynamically
allocated and thus that does not exist at compile-time.

\subsection{FLDLib Numercical Analysis Library}
\label{sec:toolchain:fldlib}


\fldlib is an open-source instrumentation library that infers accuracy
properties based on zonotopes~\cite{ghorbal09cav} over a C or C++ code. It
relies on a static analysis dedicated to verification of numerical
scenarios~\cite{goubault11vmcai} that is originally implemented in the
close-source tool \fluctuat.
\fldlib only deals with 
detecting numerical errors
and computing domains of numerical variables. Discrete values (pointers
included) are only enumerated. In particular, it has no pointer analysis. 
Therefore, it is better used
on \emph{thin scenarios} that encompass concrete test cases in small
intervals. In such scenarios, pointers have only one or two possible
value(s). This way, it is possible to get exploitable results for accuracy and
robustness properties of numerical pieces of code, even if integrated in a
larger development. 

Starting from a source code instrumented by \eacsl,
Fig.~\ref{fig:fldlib-architecture} shows how \fldlib is
integrated within a verification workflow together with \fldcompiler, and
compared to \fluctuat. Figure~\ref{fig:split-section} shows how \fldlib
handles a split-merge section generated by \fldcompiler: this section now focuses
on explaining its main parts, that are path exploration,
constraint propagation and domain computation. Even if not detailed here, \fldlib
also supports the built-ins of Fig.~\ref{fig:builtins} (after their translation by \eacsl
into their \fldlib counterparts).


\begin{figure}[tb]
{
\centering
\vspace{2mm}
  \begin{minipage}[l]{.5\textwidth}\centering\scalebox{.75}{
    \scriptsize
    \tikzstyle{flow-comp}=[ultra thick, ->, >=latex]
    \tikzstyle{flow-inte}=[dashed, ->, >=latex]
    \tikzstyle{work}=[<->, >=latex]
    \tikzstyle{element}=[draw, rectangle, text width={width("Executable code")+2pt}, align = center]
    \tikzstyle{component}=[draw, rounded rectangle, text width={width("Frama-C / Eva")+1pt}, align = center]
    \begin{tikzpicture}[node distance = 4mm]
      \node (Source) [element] {Source code} ;
      \node (O0) [below = 8mm of Source] {} ;
      \node (FramaC) [component, left = 5mm of O0] {Frama-C/Eva} ;
      \node (FLDCompiler) [component, below = of O0] {FLDCompiler} ;
      \node (O1) [below = of FLDCompiler] {} ;
      \node (Compiler) [component, below = of O1] {Compiler: g++, clang++} ;
      \node (FLDLib) [component, left = 5mm of O1] {FLDLib} ;
      \node (Toolkit) [element, fit = (FramaC) (FLDCompiler) (FLDLib) (Compiler)] {} ;
      \node (Executable) [element, below = 6mm of Compiler] {Executable code with analysis} ;
      \node (Verdict) [element, below = 6mm of Executable] {Robustness \& accuracy verdict} ;
      \node (Fluctuat) [element, right = 1mm of Toolkit] {Fluctuat} ;
      \draw [work] (FramaC.east) -| (FLDCompiler.150) ;
      \draw [work] (FLDCompiler.west) -| (FLDLib.30) ;
      \draw [work] (FLDLib.-30) |- (Compiler.west) ;
      \draw [flow-comp] (Source) -- (FLDCompiler) -- (Compiler) -- (Executable) -- (Verdict) ;
      \draw [flow-inte] (Source.east) -| (Fluctuat) |- (Verdict.east) ;
    \end{tikzpicture}
    }
    \captionof{figure}{Tool architecture.}
    \label{fig:fldlib-architecture}
  \end{minipage}
  \begin{minipage}[r]{.5\textwidth}\centering\scalebox{0.96}{
    \scriptsize
    \tikzstyle{node} = [draw, rectangle]
    \tikzstyle{expl} = [draw, rounded rectangle]
    \tikzstyle{size} = [minimum width = 2.3cm, minimum height = 0.4cm]
    \tikzstyle{link} = [thick, >=latex]
    \tikzstyle{rest} = [dashed]
    \begin{tikzpicture}[node distance = 1cm]
      \node (O) [] {} ;
      \node (SP) [node, size, right = 0.25cm of O]
        {\Split} ;
      \node (IX) [node, size, below = 0.8 cm of SP] 
        {\texttt{index=(int)in;}} ;
      \node (OU) [node, size, below = 0.8 cm of IX]
        {\texttt{out=...}} ;
      \node (MG) [node, size, below = 0.8 cm of OU]
        {\Merge} ;
      \draw [link, ->] (MG.west) -- ++(-0.25,0) |- (SP.west) ;
      \draw [link,  ->] (SP.south) -- (IX.north) ;
      \draw [link,  ->] (OU.south) -- (MG.north) ;
      \draw [link, <-] (OU.170) -- ++(0,0.5) -- (IX.195) ;
      \node (SP-Ex) [expl, size, right = 0.65cm of SP]
        {path selection} ;
      \node (E) [below = 0.9cm of SP] {} ;
      \node (CT) [expl, text width = 1.8cm, align = center, right = 1.68cm of E]
        {constraint propagation} ;
      \node (OU-Ex) [expl, size, right = 0.65cm of OU]
        {domain computation} ;
      \node (MG-Ex) [expl, size, right = 0.65cm of MG]
        {union} ;
      \node (ACC-SP1) [above left = 0.3cm and 0.43cm of SP-Ex.north west] {} ;
      \node (ACC-SP2) [below left = 0.3cm and 0.43cm of SP-Ex.south west] {} ;
      \draw [decorate,decoration={brace}] (ACC-SP1) -- (ACC-SP2) ;
      \node (ACC-CT1) [above left = 0.15cm and 0.53cm of CT.north west] {} ;
      \node (ACC-CT2) [below left = 0.15cm and 0.53cm of CT.south west] {} ;
      \draw [decorate,decoration={brace}] (ACC-CT1) -- (ACC-CT2) ;
      \node (ACC-OU1) [above left = 0.3cm and 0.43cm of OU-Ex.north west] {} ;
      \node (ACC-OU2) [below left = 0.3cm and 0.43cm of OU-Ex.south west] {} ;
      \draw [decorate,decoration={brace}] (ACC-OU1) -- (ACC-OU2) ;
      \node (ACC-MG1) [above left = 0.3cm and 0.43cm of MG-Ex.north west] {} ;
      \node (ACC-MG2) [below left = 0.3cm and 0.43cm of MG-Ex.south west] {} ;
      \draw [decorate,decoration={brace}] (ACC-MG1) -- (ACC-MG2) ;
      \draw [link, <-] (SP.north) -- ++(0, 0.4) ;
      \draw [link, ->] (MG.south) -- ++(0,-0.4) ;
      \draw [rest] (IX.-15) -- ++( -55:0.40) ;
      \draw [rest] (IX.-27) -- ++( -65:0.30) ;
      \draw [rest] (IX.-59) -- ++( -80:0.28) ;
      \draw [rest] (IX.239) -- ++(-100:0.28) ;
      \draw [rest] (IX.207) -- ++(-115:0.30) ;
    \end{tikzpicture}
    }
    \captionof{figure}{Handling a split-merge section.}
    \label{fig:split-section}
  \end{minipage}
}
\end{figure}

\begin{figure}[tb]
	\includegraphics[width=\textwidth]{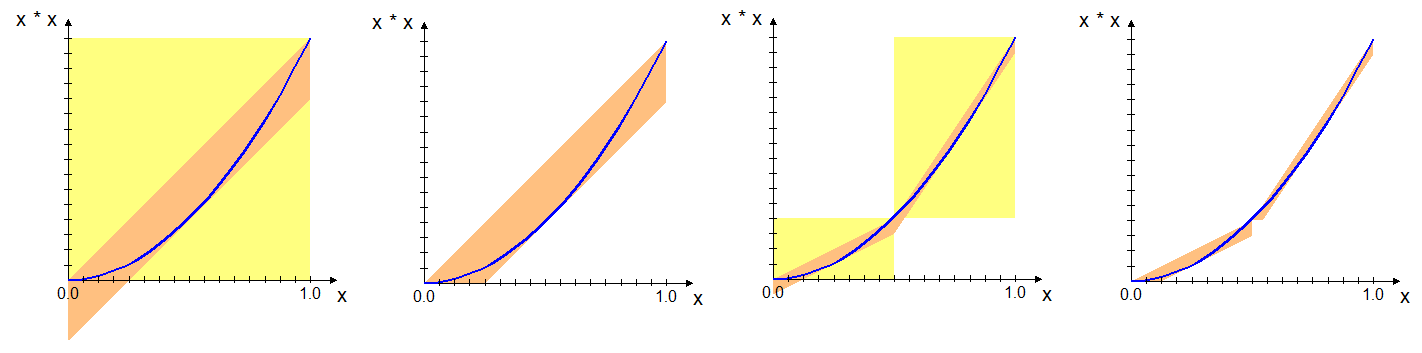}
	\vspace{-3mm}
	\caption{Function $x^2$ 
		abstracted (a) with intervals
		(yellow) and affine forms (orange) shown separately, and (b) the resulting intersection.
		The same abstractions with a subdivision, 
		(c) shown separately, and (d) the resulting intersection.}
	\label{fig:zonotope-interval}
\end{figure}

\paragraph{Domain computation.}

\fldlib domains combine intervals and \emph{zonotopes}~\cite{ghorbal09cav}. Zonotopes allow to
    maintain linear relationships between program variables $V$ that share the same
    perturbations (noise symbols) by mapping $V$ to affine forms. 
    Sharing noise symbols between variables helps at keeping precise information since it means
    that the source of uncertainty is the same. 
    We do not detail the zonotope domain here for lack of space, 
    but only illustrate by examples (see Fig.~\ref{fig:zonotope-interval})
    the benefits of combining zonotopes and intervals, in particular, 
    with a domain subdivision.
    For instance, if $x \in [0,1]$,
    an interval is more precise than a zonotope for representing $x^2$
    (providing an interval $x^2\in[0, 1]$ instead of $[-0.25,1]$, cf. the
    projection of abstractions onto the $x \times x$ axis in
    Fig.~\ref{fig:zonotope-interval}a), but less precise for 
    representing $x-x^2$ ($[-1,1]$ instead of $[0, 0.25]$, cf. the distance from the diagonal in Fig.~\ref{fig:zonotope-interval}a).
    The intersection of both abstractions
    provides more precise results (Fig.~\ref{fig:zonotope-interval}b).    
    A subdivision of the input interval into two intervals significantly
    improves the results (Fig.~\ref{fig:zonotope-interval}c,d).
    The union operation mentioned in Fig.~\ref{fig:split-section} does not try to keep 
    linear relationships coming from zonotopes but ensures the preservation of intervals.

The main principle of \fldlib consists in
redefining \texttt{double} and
\texttt{float} types by a structure containing four fields representing the abstraction: (1) the
floating point domain \textit{float} as an interval, (2) the real-number domain \textit{real} as a zonotope,
(3) the error domain \textit{err} as a zonotope,
and (4) the relative error defined by an
interval. Like \cadna~\cite{FJ-JMC-CPC-2008}, \fldlib uses C++ operator
overloading to propagate these four domains over the program execution by
redefining comparisons and casts from floating-point types to integral
types. Operator overloading 
is particularly interesting here since it limits necessary
source-to-source transformation. 
A similar approach could be applied to \C programs with no
operator overloading capabilities, where such a transformation can be automatically
done by the \name{Clang} compiler. We have done promising initial experiments on Ada
programs that also support operator overloading through the \name{libadalang}
library. 

\algdef{SE}[DOWHILE]{Do}{doWhile}{\algorithmicdo}[1]{\algorithmicwhile\ #1}%
\renewcommand{\algorithmiccomment}[1]{{\hfill\color{blue}{$\triangleright$ #1}}}

\begin{figure}[tb]
\begin{algorithmic}[1]
\State \Comment{Start of Split}
\State $\textit{stack}  \gets \textit{push}(\textit{stack},
  \textit{current-path-explorer}, \textit{end-explore})$
\!\!\!\!\Comment{save outer section context}
\State $\textit{current-path-explorer} \gets\newline ~\qquad 
  \textit{new\_path\_explorer\_from\_mode}(\textit{current-path-explorer},
  \textit{id})$ 
\Comment{create a new one}
\State $\textit{end-explore} \gets \textit{end-explore\_id}$ 
\Comment{define label to explore next path of section \textit{id}}
\State $M_\Sigma \gets \emptyset$ \Comment{merged memory $M_\Sigma$ is initially empty}
\State $M_\mathrm{saved} \gets
  \textit{create\_local\_memory\_for}(\textit{M},
  \textit{save-list})$ \Comment{save cur. memory for \textit{save-list}}
\Do \Comment{iterate over execution paths}
\State $\textit{M} \gets \textit{restore\_memory}(M_\mathrm{saved},
  \textit{save-list}, \textit{M})$ \Comment{restore  $M_\mathrm{saved}$ to $M$ for \textit{save-list}}
\State \Comment{End of Split}
\State \framebox{instructions of the section}
\Comment{intrumented to ensure exploration of paths}  
\State \Comment{Start of Merge}
\State $\textit{M} \gets \textit{merge\_unstable}
  (\textit{current-path-explorer}, \textit{merge-list}, \textit{M})$
\Comment{merge path data}    
\State\unskip \emph{end-explore\_id}:\Comment{jump here to explore the next path}
\State $M_\Sigma \gets \textit{merge}(M_\Sigma,
  \textit{merge-list}, \textit{M})$ \Comment{merge domains for \textit{merge-list} into $M_\Sigma$}
\doWhile {$\textit{next\_path}
  (\textit{current-path-explorer})$} \Comment{choose next path if any, or exit loop}
\State $\textit{M} \gets M_\Sigma$ \Comment{set current memory $M$ to merged memory $M_\Sigma$}
\State $\textit{current-path-explorer}, \textit{end-explore}
  \gets \textit{pop}(\textit{stack})$
\Comment{get the outer section's context}  
\If {$M_\Sigma = \emptyset$} \Comment{if no feasible path in the section}
\State $\textit{\bf goto } \textit{end-explore}$
\Comment{then explore next path in the outer section}
\State \Comment{End of Merge}
\EndIf
\end{algorithmic}
\vspace{-3mm}
\caption{Instrumentation of a split-merge section.}
\label{fig:sm-sec-instrumented}
\end{figure}

\paragraph{Constraint propagation.}

Comparison operators and casts from floating-point values to integer values are
likely to introduce constraints on noise symbols $\epsilon_i$. 
Consider for instance a conditional whose guard is $x \geq 0$ for some variable
$x$ whose abstraction is $\textit{real} = \epsilon_0$, $\textit{err} =
10^{-7}\epsilon_1$, and $\textit{float} = [-1, 1]$. It may lead to six execution
flows depending on the evaluation of the guard as a float (denoted $c_f$) or as
a real (denoted $c_r$): two if $c_f$ and $c_r$ are evaluated to the same truth
value, and four if these values diverge and the execution after the conditional
is interpreted either as a floating-point or a real number ($2 + 2 \times 2 = 6$). Choosing an
execution flow among these six ones is called a \emph{local decision}.


On our example, if $\texttt{x}\idfloat$ and $\texttt{x}\idreal$ are both
positive, then $\epsilon_0 \in [0,
1]$. Therefore, it defines a new noise symbol $\epsilon_d \stackrel{\rm def}{=}
2\epsilon_0 - 1$ (so $\epsilon_0 = 0.5 \epsilon_d + 0.5$) , constrained to be in
the interval $[-1, 1]$. Every affine form 
containing the shared symbol $\epsilon_0$ is then notified. If the replacement
of $\epsilon_0$ by $0.5 + 0.5\epsilon_d$
would improve the amplitude of the affine form by 5\%, this replacement becomes
effective even if it may lose some relationships with other
variables. This rate of 5\% is proved to be good in practice.
In this particular example, $x$ becomes $\textit{real} = 0.5 + 0.5\epsilon_d$,
$\textit{err} = 10^{-7}\epsilon_1$, $\textit{float} = [0, 1]$.
Similarly, the unstable flow $\texttt{x}\idfloat \geq 0, \texttt{x}\idreal < 0,
0 < \texttt{x}\iderror$ requires $\epsilon_1 \in [0, 10^{-7}]$, $\epsilon_0
\in [-10^{-7}, 0]$, since $\textit{float} = \textit{real} +
10^{-7}\epsilon_1 \geq 0$. After replacement, $x$ becomes $\textit{real} =
-5\times10^{-8} + 5\times10^{-8}\epsilon_{d0}$,
$\textit{err} = 5\times10^{-8} + 5\times10^{-8}\epsilon_{d1}$, $\textit{float} =
[0, 10^{-7}]$.

\paragraph{Path exploration within split-merge sections.}

\fldlib instruments the constructs 
$\Split(\textit{id}, \textit{save-list})$ and 
$\Merge(\textit{id}, \textit{merge-list})$
generated by \fldcompiler, where \textit{id}  is a unique identifier for each section. 
As we outlined by Fig.~\ref{fig:split-section},
they encompass the instructions of the section. We detail this in Fig.~\ref{fig:sm-sec-instrumented}.
Split-merge sections can be nested.
As explained in 
Section~\ref{sec:toolchain:fldcompiler}, the current memory state $M$ is saved (for the variables of \textit{save-list}) in $M_\mathrm{saved}$ and restored back into $M$
to start every new execution of the section
from the same state (cf. lines 6, 8 in Fig.~\ref{fig:sm-sec-instrumented}). 

The outer section context is saved in a stack at line 2 and retrieved at line 17. 
The label \textit{end-explore} defines a label where the execution jumps to explore a next path if an inner split-merge section found no feasible path, or if an infeasible path is encountered in the current section while exploring the instructions at line 10. The jump at line 19 shows the first usage. If feasible paths are found, the execution continues with a merged state (line 16)
after retrieving the context (line 17).

Another stack,  \textit{current-path-explorer}, ensures the storage and exploration of  
paths. The call to \textit{next\_path} at line 15 
chooses the next path to be executed, if any, and stores it in
\textit{current-path-explorer},
or exits the loop otherwise. 
The exploration is performed in a depth-first search and relies on an instrumentation of
comparison operations and float-to-interger casts encountered in the section.
The instrumentation is realized by overloading comparison and cast operations in \Cpp.

\lstset{basicstyle=\normalsize\ttfamily}
In practice, in addition to the \Split/\Merge directives, \fldlib adds at the
beginning of each source file a new header \lstinline{float_diagnosis.h}. This
header replaces the definition of \texttt{double} and \texttt{float} types by
the representation of their abstract domains (as explained earlier in this
section). 
The implementation of \textbf{goto} \textit{end-explore} benefits from the \Cpp exception
mechanism. The \texttt{main} function is augmented with a header and a footer
in order to initialize and clean up the analysis state.

Last but not least, \fldlib prevents \textit{merge-list} from
containing \textit{int} variables in order to keep precise relationships between
discrete and continuous variables. These constraints may move forward
some \Merge constructions until the \textit{int} variables become useless.


%
%
\section{Experimental Results}
\label{sec:experimental}

The toolchain presented in this paper has been evaluated on a benchmark of
small-size \C examples and experimented on two industrial case studies.

\paragraph{Benchmarks.}

We use the benchmarks from~\cite{goubault13aplas}. It contains several
small-size \C examples that may be categorized as follows.
\begin{description}
  \item{\bf Simple examples}
    show basic computations that focus on accuracy properties.
\item{\bf
    Unstable branches} are robustness tests for unstable branch handling.
\item{\bf Interpolation tables} contain various ways to
    compute an interpolation table
    . They also focus on testing robustness of unstable branches.
\item{\bf Maths} models functions of \texttt{math.h} for error
    estimation.
%
\item{\bf Miscellaneous} contain various other examples. File \texttt{filter.c}
    is a second order linear filter that focuses accuracy
%
    File \texttt{patriot.c} is an historical example that contains a sum of 0.1
    whose error shifts over time.
%
File \texttt{complex\_LU.c} finds a
    vector $X$ such that $M(X) = (Y)$ for a square matrix $M$ with a
    Lower/Upper decomposition.
%
    File \texttt{complex\_intersect.c} shows iterative computations.
%
    File \texttt{scanf.c} show how to manage external library
    functions not related to floating-point operations.
\end{description}

Each example have been annotated with \acsl assertions modeling the expected
properties in order to use our toolchain. All of them have also been run
with \fluctuat~\cite{goubault11vmcai},
\name{Precisa}~\cite{DBLP:conf/lopstr/TitoloMFM18}
and \name{FpDebug}~\cite{benz12pldi}. Figure~\ref{fig:comparison-table} presents
the results of our evaluation.
\texttt{ko} identifies a case where the tool is not pertinent (scalability,
unmanaged unstable branch) or when a reasonable bound for accuracy is not
verified. \texttt{n/t}
means ``not translated'' into \name{PVS} for \name{Precisa}. A number reports
the accuracy that the tool infers. If written in bold, that is the best
accuracy for a particular example. Therefore, the table clearly shows
that \textbf{our toolchain has almost always the best accuracy.}


\begin{figure}[tb]\centering
\noindent{\scriptsize
\begin{tabular}{lclclclcl}
\midrule
 \textbf{file/variable}  & ~ & \textbf{Our toolchain}      & ~ & \fluctuat  & ~ & \name{Precisa} & ~ & \name{FpDebug} \\
\midrule
\textbf{Simple examples:} \\
 absorption.c/z          &   & {\bf 1e-8}               &   & {\bf 1e-8}         &   & 5.96e-8          &   & 1e-8             \\
 associativity.c/u       &   & {\bf 6.67e-16}           &   & 1.55e-15           &   & 4.21e-15         &   & -2.22e-16        \\
 division.c/z2           &   & {\bf 1.805e-16}          &   & 5.55e-16           &   & 5.55e-16         &   & -1.57e-17        \\
 exp.c/y                 &   & {\bf 4.47e-13}           &   & 5.61e-13           &   & 4.45e-12         &   & ko               \\
 polynome.c/t            &   & 3.576e-06                &   & {\bf 2.57e-06}     &   & 9.67e-06         &   & 2.15e-07         \\
 relative.c/z            &   & {\bf 2.32e-12}           &   & {\bf 2.32e-12}     &   & 6.59e-12         &   & 1.82e-13         \\
 triangle.c/A            &   & {\bf 1.44e-04}           &   & 1.73e-04
 &   & 8.09e-04         &   & 0.0              \\
\midrule 
\textbf{Unstable branches:} \\
 comp\_abs.c/z           &   & {\bf 3.58e-07}           &   & 2 (false alarm)    &   & 4 (false alarm)  &   & -2.85e-08        \\
 comp\_cont.c/z          &   & {\bf 5.03e-05}           &   & 9.03e-05           &   & 3 (false alarm)  &   & -2.25e-08        \\
 comp\_cont\_nested.c/z  &   & {\bf 1.0e-18}            &   & {\bf 1.0e-18}      &   & n/t              &   & -1.0e-18         \\
 comp\_cont\_mult.c/res  &   & {\bf 3.40e-05}           &   & 105 (false alarm)  &   & 192              &   & ko (unstable)    \\
 comp\_disc\_nested.c/z  &   & {\bf 0.1} (true alarm)   &   & {\bf 0.1}          &   & n/t              &   & ko               \\
 comp\_disc.c/z          &   & {\bf 0.5} (true alarm)   &   & 1.0                &   & 2                &   & ko               \\
 comp\_model\_err.c/S    &   & {\bf 0.023} (true alarm) &   & 3.82e-01
 &   & ko               &   & ko               \\
 \midrule
\textbf{Interpolation tables:}\\
 inter\_cond.c/res       &   & {\bf 1.33e-05}           &   & 105 (false alarm)  &   & 191              &   & 4.77e-07         \\
 inter\_loop.c/res       &   & {\bf 1.45e-06}           &   & 4.05e-06           &   & 33               &   & -4.60e-07        \\
 inter\_tbl\_cast.c/out1 &   & {\bf 4e-06}              &   & 77.1 (false alarm) &   & time out         &   & -1.04e-15        \\
 inter\_tbl\_loop.c/res  &   & {\bf 4e-08}              &   & ko                 &   & n/t              &   & -1.04e-15        \\
 motiv\_example.c/out1   &   & {\bf 1.19e-07}           &   & 77.1 (false alarm) &   & time out         &   & -1.04e-08        \\
 motiv\_example.c/out2   &   & {\bf 4} (true alarm)     &   & 95.1
 &   & time out         &   & ko               \\
 \midrule
 \textbf{Maths:}\\
 sin\_model\_error.c/res &   & {\bf 2.57e-16}           &   & {\bf 2.57e-16}     &   & n/t              &   & 8.79e-18         \\
 sqrt\_unroll.c/t.v      &   & {\bf 7.10e-15}           &   & 7.81e-14           &   & n/t              &   & -4.81e-15        \\
 sqrt\_fixpoint.c/res    &   & {\bf 3.15e-15}           &   & 1.39e-14           &   & n/t              &   & 3.51e-16         \\
 \midrule
\textbf{Miscellaneous:}\\
 filter.c/S              &   & {\bf 1.56e-14}           &   & 1.65e-14
 &   & time out         &   & 1.44e-16         \\
 patriot.c/t             &   & {\bf 1.91e-04}           &   & {\bf 1.91e-04}     &   & time out         &   & {\bf 1.91e-04}   \\
 complex\_LU.c/det       &   & {\bf 7.14e-15}           &   & ko                 &   & ko               &   & {\bf 7.14e-15}   \\
 complex\_intersect.c/x  &   & {\bf 5.32e-01}           &   & 0.2                &   & ko               &   & ko               \\
 scanf.c/res             &   & {\bf 5.96e-07}           &   & ko                 &   & ko               &   & ko               \\
\midrule
\end{tabular}
}
\caption{Tool comparison over small-size \C examples.}
\label{fig:comparison-table}
\end{figure}

Since \fldlib uses the same reasoning as \fluctuat except on the constraint
management, many results are merely the same. However, \fluctuat has only a
minor support of unstable branches. \name{FpDebug} also lacks of support of
unstable branches. That explains why our toolchain is often better than these
tools.
The comparison with \name{Precisa} is somehow biased since \name{SMT}
optimization with \name{FPRock} was not activated. We do not know if it would
have scaled better with this optimization.  Nevertheless our toolchain aims to
provide guaranteed accuracy analysis with unstable branches on existing \C code
containing loops and several thousands of lines of source code,
while \name{Precisa} is more concerned with robustness proofs of smaller
algorithms.

\paragraph{Industrial case studies.}

We also experimented our toolchain on two (non public) industrial case studies
(synchronous reactive systems of several dozens of thousands of lines of code).
The \C code of the first one was automatically generated, whereas the code of
the second one was written by hand in \Cpp. The first one contains computations
that represent physical models, with many components like interpolation tables.
One component (convergent linear filters) deals with memory management and
not only with numerical computations. The second use cases contains solving
algorithms and uses external libraries base on templates. 


\fldcompiler was used to identify the location of the \Split{}-\Merge{}
annotations and the list of variables that must be saved and merged. Even if we
only used its syntactic 
version that is not based on the \Eva plug-in of \framac (resulting in a loss of
precision), the results were pretty good and useful. However, these case studies
shown that \fldcompiler need to be extended to provide better results on some
linear algebra algorithms and some discontinuous unstable branches. For example,
the determinant computation is a continuous formula but it often internally uses
a LU (Lower/Upper) matrix decomposition that contains many unstable
branches due to the choice of the best pivoting number. Delaying the merging
point as done by \fldcompiler results in that case in a combinatorial
explosion. In practice, 
we choose only one particular matrix decomposition to go on: by default it is
the one that has been created by the last execution path between the split and
the merge.

\fldlib scales better that \name{Fluctuat} on these case studies since it does
not care about pointers. 
Nevertheless, \fldlib' scalability is directly related to the trade-off between
precise results and analysis' times: if the number of noise symbols in
zonotopes is not bounded, the analysis may be quadratic. In practice, an option
sets a bound to limit the number of noise symbols introduced in an affine form.

On the first generated industrial C code, our toolchain succeeds in keeping a
reasonable error for a thin scenario and such avoiding excessive
over-approximations.
On the second industrial C++ code, the guaranteed numerical error
delivered by \fldlib increases at every loop cycle
and false alarms appear because more and more unstable branches are
detected. That leads to a combinatorial explosions in solving algorithms
of \fldcompiler (when the merge point is far from the split point in terms of
instructions). In such a case \fldlib has rather
been used to identify the tricky numerical parts of a big code.

All in all, these industrial use cases demonstrate that \textbf{our toolchain
scales on thin scenarios} up to several dozens of thousands of lines of code. At
worst, a few \Split{}-\Merge{} points have to be manually set and \fldlib
provides an helpful support for this task.
It is also worth noting that \fldlib can be replaced by \name{Cadna} to obtain
a stochastic analysis that scales better, even if the results are
non-necessarily sound but close to the expected ones. 
We have also experimented the exact part of \fldlib that roughly works
like \name{FpDebug}, but at source code level. With this component we obtain
under-approximated results.

Last but not least, our toolchain may be easily integrated within a continuous
integration setting. For that purpose, it only requires to instrument the unit
test files. Any other file (including library file) can remain unchanged.






%
\section{Related Work}
\label{sec:related-work}
Verification of accuracy properties remains a challenging
research topic.
During the last fifteen years, many tools and academic prototypes have been
developed to tackle this problem.
They can be classified in two categories: \emph{testing} tools and \emph{static
analysis} tools.

Let us consider testing tools first.
\name{FpDebug}~\cite{benz12pldi} and \name{Herbgrind}~\cite{sanchez-stern17corr}
are two tools based on \name{Valgrind}~\cite{nethercote07jun} that are designed
to avoid false positive reports.
\name{FpDebug} relies on \name{MPFR}~\footnote{\url{https://www.mpfr.org}} to
associate a 
highly-precise value to each floating-point value of the tested program.
If the difference between both values is too large, an error is reported.
It guarantees that no reported problem is a false positive, but it can miss
problems that need better precision.
\name{Herbgrind} evaluates rounding errors in a similar way but, in addition, it
uses symbolic execution to detect important accuracy loss and track the root
causes of those losses.
Unlike our toolchain, \name{FpDebug} and \name{Herbgrind} avoid false positive
reports and scale up on bigger programs.
However, our toolchain can guarantee that there is no error in the program.
\name{Verrou}~\cite{DBLP:conf/cav/FevotteL17},
\name{Cadna}~\cite{FJ-JMC-CPC-2008}, and \name{Verificarlo}~\cite{verificarlo}
are other examples of testing tools that aim at reporting possible instances of errors.
They use stochastic protocols to give error estimations.
The core idea consists in randomly (with a selected probability) changing
the rounding mode used for each
floating-point operation encountered during the program execution.
For each execution, the obtained floating-point values differ, and with enough
executions, an estimation can be made. Note that \name{Cadna} infers three
different executions during one analysis.
The stochastic process allows to obtain a good confidence in such an estimation,
even if it cannot provide any guarantee.
Like our toolchain, those three tools do not avoid false positive reports
because of the stochastic process.
However, unlike us, they cannot guarantee the absence of errors for the same
reason.

For static analysis tools, different trade-offs exist between scalability and
tightness of over-approximations.
\name{Precisa}~\cite{DBLP:conf/lopstr/TitoloMFM18} and
\fluctuat~\cite{goubault11vmcai} are based on abstract interpretation and
favor scalability.
Both of them handle unstable tests; \name{Precisa} manage them more rigourously,
but Fluctuat scales reasonably for
programs of a few thousand lines of code. \name{Precisa} uses interval
arithmetic combined with branch-and-bound optimization and symbolic error
computations, while \fluctuat relies on the zonotope abstract
domain~\cite{ghorbal09cav} to represent both the values and the errors along the
analyzed program.
Compared to \name{Precisa} and \name{Fluctuat}, our toolchain scales better and
can handle I/O and memory manipulations without the need of stubs.
Another tool is \name{FPTaylor}~\cite{DBLP:journals/toplas/SolovyevBBJRG19}. It
favors tightness. For that purpose, it formulates the problem of bounding errors
as an optimization problem that is soundly solved by using first-order Taylor
approximations of arithmetic expressions.
Bounds are found on the transformed expressions by using expensive optimization
procedures. However, they allow \name{FPTaylor} to provide tight over-approximations of absolute and relative errors.
\name{FPTaylor} generally provides tighter approximations than our toolchain.
However, unlike our toolchain, it cannot analyze large programs and handles
neither loops, nor I/O operations, nor unstable tests.
Finally, \name{Gappa}~\cite{DBLP:journals/toms/DaumasM10} presents a third
possible trade-off.
Indeed, \name{Gappa} is intended to help verifying and formally proving
properties on numerical programs.
It is based on interval arithmetic and several rewriting rules for
floating-point rounding errors expressions.
It can generate proof certificates that can be verified using the \name{Coq}
proof assistant and is also used as a backend prover for the \name{Why3}
software verification platform.
The interest of our toolchain lies in its ability to relatively scale and
in its management of I/O and memory manipulations without the need of stubs.
However, the goal of those tools is to prove numerical properties for large input
domains, while our toolchain is limited to thin scenarios.

\section{Conclusion and Perspectives}
\label{sec:conclusion}
Assessment of numerical accuracy in critical programs is crucial to prevent
accumulation of 
rounding errors that can provoke dangerous bugs.
In this work, we presented the design and implementation of a new toolchain for evaluation of numerical accuracy properties in C programs. 
It is based on abstract compilation, embedding an accuracy analysis 
engine into the compiled code whose execution allows accuracy property assessment. 
This work relies on several independent original contributions: 
an extension of the runtime 
verification tool \eacsl to precisely support rational numbers, 
the design of the abstract compiler \fldcompiler and 
an instrumentation library, \fldlib, for reasoning about numerical properties.

The proposed approach combines the precision of testing-based techniques with
the capacity of static analysis to analyze several executions at the same time.
Initial experiments confirm the interest of the proposed technique, in
particular, for industrial software.  Future work includes a proof of soundness
for the whole toolchain, a large evaluation on
real-life programs and an extension of the toolchain to support
all features of the C programming language, as well as 
specific \acsl built-ins (e.g. \texttt{\bs{}exact}, \texttt{\bs{}round\_error})
capable to express powerful properties over 
floating-point and real numbers.

\paragraph{Acknowledgement.} The authors thank Romain Soulat and Thales Research \& Technology for providing 
case studies and participation in the evaluation. 
We are also very grateful to Jean Gassino, Gregory de la Grange and IRSN 
for their support and evaluation on additional case studies.
\vfill\null\newpage
\bibliographystyle{splncs}
\bibliography{main}

\end{document}